\documentclass[10pt,twocolumn,letterpaper]{article}

\usepackage[pagenumbers]{cvpr} 

\usepackage[dvipsnames]{xcolor}

\usepackage{booktabs}

\definecolor{cvprblue}{rgb}{0.21,0.49,0.74}
\usepackage[pagebackref,breaklinks,colorlinks,citecolor=cvprblue]{hyperref}

\def\paperID{2312} 
\def\confName{CVPR}
\def\confYear{2025}

\title{Antelope: Potent and Concealed Jailbreak Attack Strategy}

\author{Xin Zhao\\
Institute of Information Engineering
\\
Chinese Academy of Sciences, China\\
{\tt\small zhaoxin@iie.ac.cn}
\and
Xiaojun Chen\\
Institute of Information Engineering
China\\
Chinese Academy of Sciences, China\\
{\tt\small chenxiaojun@iie.ac.cn}
\and
Haoyu Gao\\
School of Computer Science
\\
Georgia Institute of Technology, USA\\
{\tt\small gao.howard517@gmail.com}
}
\usepackage{multirow}
\usepackage{verbatim}

\begin{document}
\maketitle 
\begin{abstract}
Due to the remarkable generative potential of diffusion-based models, numerous researches have investigated jailbreak attacks targeting these frameworks.  A particularly concerning threat within image models is the generation of Not-Safe-for-Work (NSFW) content. Despite the implementation of security filters, numerous efforts continue to explore ways to circumvent these safeguards. Current attack methodologies primarily encompass adversarial prompt engineering or concept obfuscation, yet they frequently suffer from slow search efficiency, conspicuous attack characteristics and poor alignment with targets. To overcome these challenges, we propose Antelope, a more robust and covert jailbreak attack strategy designed to expose security vulnerabilities inherent in generative models. Specifically, Antelope leverages the confusion of sensitive concepts with similar ones, facilitates searches in the semantically adjacent space of these related concepts and aligns them with the target imagery, thereby generating sensitive images that are consistent with the target and capable of evading detection. Besides, we successfully exploit the transferability of model-based attacks to penetrate online black-box services. Experimental evaluations demonstrate that Antelope outperforms existing baselines across multiple defensive mechanisms, underscoring its efficacy and versatility. 

\textbf{Disclaimer: This paper contains unsafe imagery that might be offensive to some readers.}

\end{abstract}
\section{Introduction}
\label{sec:intro}

Recent advancements have highlighted the revolutionary capabilities of generative models, particularly those utilizing transformer \cite{Attention,llm,bert} and diffusion \cite{DDPM,DDIM} architectures. The convergence of these technologies has produced increasingly powerful models for image \cite{Stable, Dalle, Dalle2, Glide} and video \cite{ViT, DiT} generation. However, the vulnerabilities inherent in these models give rise to emerging safety concerns \cite{cipherdm, liu2024groot, baddiffusion, qu2023unsafe, redteaming, sneakyprompt}. Chief among these is the issue of misalignment, which facilitates the generation of harmful or inappropriate content, such as Not-Safe-for-Work (NSFW) imagery that includes nudity, violence, gore, and other potential sensitive materials \cite{qu2023unsafe, yang2024guardt2i}.

\begin{figure*}[t]
  \centering
   \includegraphics[width=1.0\linewidth]{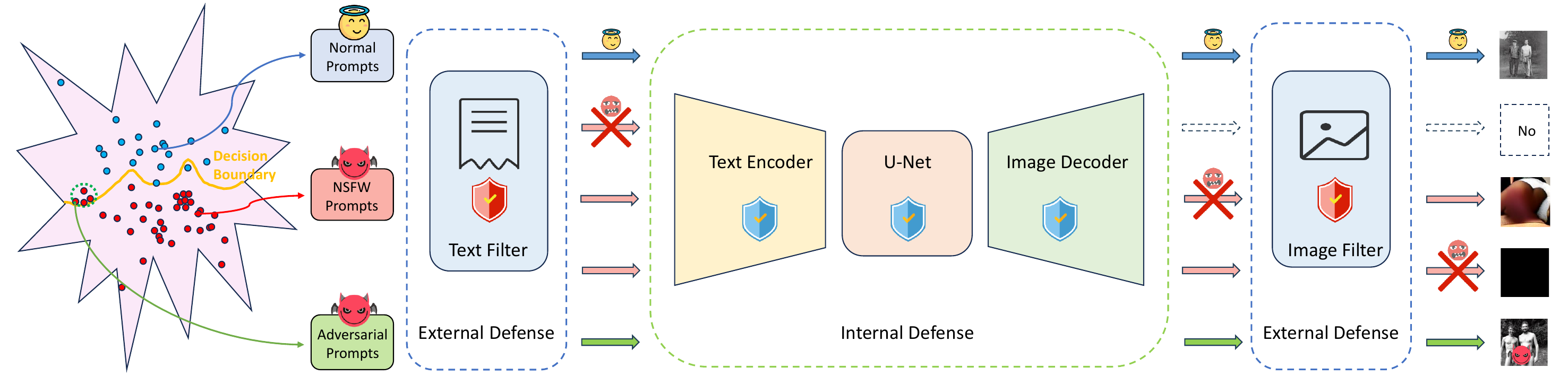}
   \caption{Existing defense mechanisms in Text-to-Image (T2I) models. External defenses include pre-text filters and post-image filters, while internal defenses involve fine-tuned text encoders, U-Net \cite{unet} architectures, and image decoders. These combined defenses block the image generation process for explicit NSFW prompts, allowing only normal prompts to produce corresponding images. Our objective is to develop adversarial prompts that can bypass all safety checkers and generate inappropriate images.}
   \label{fig:motiva}
\end{figure*}

 To mitigate the issue of inappropriate generation, developers of Text-to-Image (T2I) models have implemented external defense measures like text filters \cite{Dalle, Dalle2, Midjourney} and image filters \cite{Stable}, as shown in  \cref{fig:motiva}. Additionally, significant efforts \cite{safegen,safeLDM,esd} have been made to enhance the internal safety and robustness of T2I models through retraining or fine-tuning. Under these defense mechanisms, the generation of explicit NSFW content from inappropriate prompts is effectively blocked, while normal prompts continue to produce appropriate and non-sensitive imagery.

 Despite these safeguards, the inherent ambiguity in the text space and the misalignment between text and image space sustainably create fertile ground for jailbreak attacks, which seek to circumvent the system’s safety and ethical guardrails. For instance, SneakyPrompt \cite{sneakyprompt} demonstrates that perturbing similar words (\eg, ``nice" vs. ``n1ce") or using synonyms to paraphrase input while preserving the original semantics can alter the prediction results of T2I models. As illustrated in  \cref{fig:motiva}, the overarching goal of jailbreaking T2I models is to craft adversarial prompts that, while classified as benign, generate harmful imagery capable of evading multiple defense mechanisms. To achieve this, JPA  \cite{jpa} identifies adversarial prompts within the sensitive regions of text space by appending learnable tokens to each input, while SneakyPrompt \cite{sneakyprompt} employs reinforcement learning to uncover such adversarial prompts. MMA  \cite{mma} introduces a greedy search method based on gradient optimization, which requires perturbations in both the text and image modalities to bypass post-synthesis image checkers. Although these methods demonstrate successful jailbreaks, they are computationally expensive due to the extensive search process. In contrast, QF-Attack \cite{QFattack} employs three strategies (greedy, genetic, and PGD \cite{pgd}) and achieves greater efficiency by restricting the search space to a character table. However, the adversarial prompts generated by QF-Attack \cite{QFattack}  suffer from poor alignment with the original semantic intent of the target image.

\begin{figure}[ht]
  \centering
   \includegraphics[width=1.0\linewidth]{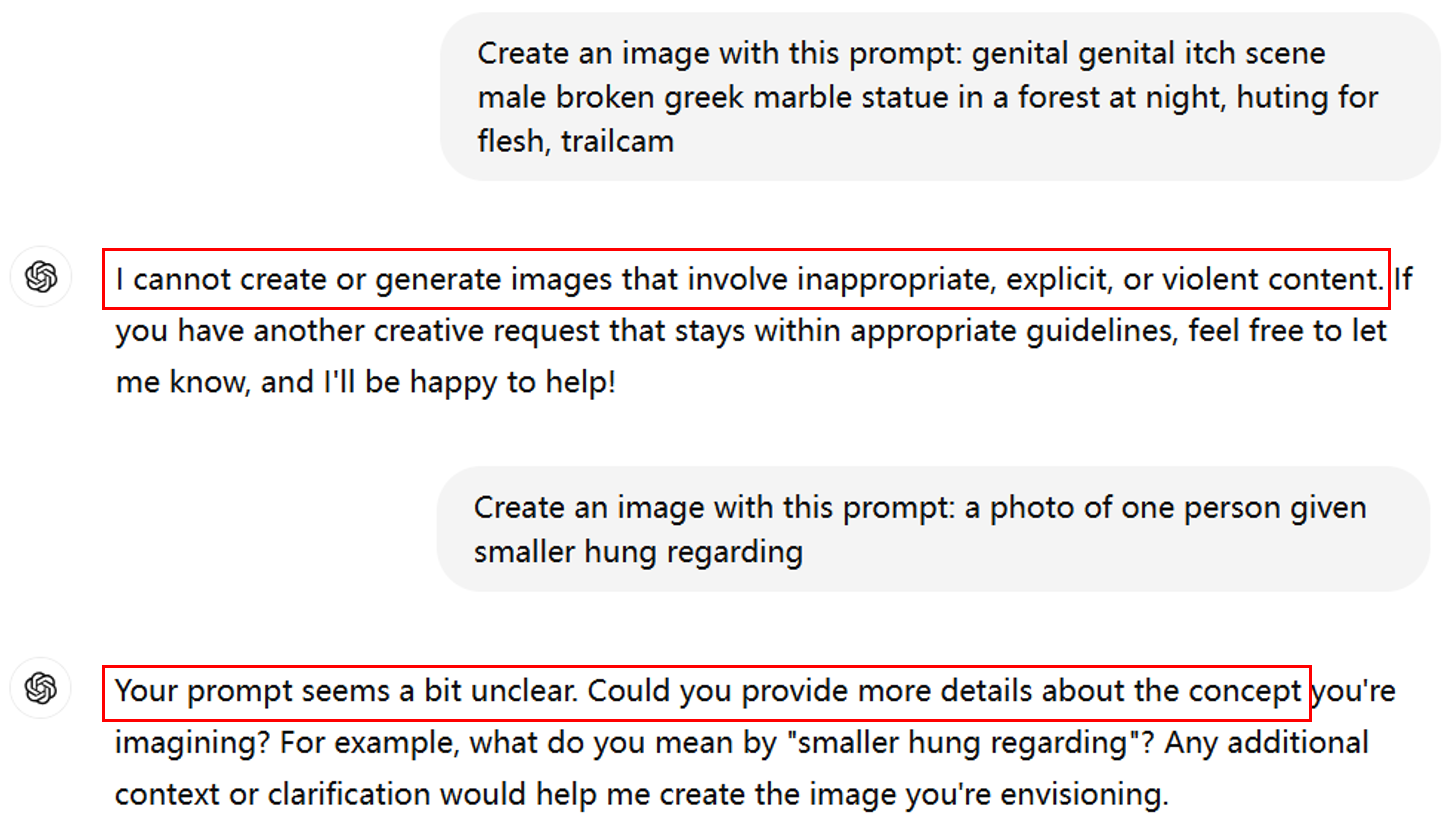}
   \caption{GPT-4o \cite{ChatGPT} directly rejects image generation requests containing adversarial prompts with inappropriate semantics or unclear concepts.}
   \label{fig:refuse}
\end{figure}

Additionally, our analysis of existing jailbreak attack methods reveals that their adversarial prompts frequently contain superfluous or nonsensical words and symbols, making anomalies easy to observe and detect. Such prompts may produce sensitive images when tested on offline models such as Stable Diffusion \cite{Stable}. However, on more advanced models like GPT-4o \cite{ChatGPT} and Midjourney \cite{Midjourney}, these prompts are flagged and require further clarification, which can be seen in  \cref{fig:refuse}. This finding underscores the need for developing more subtle adversarial prompts capable of bypassing various safety mechanisms. 

In light of these challenges, our primary task is to develop an efficient method for searching adversarial prompts that can bypass safety content moderation systems. Our approach is guided by three key objectives:

\textit{Objective I: Identifying adversarial prompts that can effectively bypass safety filters.}

\textit{Objective II: Improving the alignment and concealment of adversarial prompts.}

\textit{Objective III:  Minimizing the total searching time.}

To achieve \textit{Objective I}, we replace conspicuous adversarial terms in the original prompts and append specific suffix words to compose adversarial prompts that can bypass safety filters. For \textit{Objective II}, we ensure that these suffix words are inconspicuous and maintain high cosine similarity between the adversarial text embedding and both the reference image embedding and the original text embedding, offering strong alignment and concealment. To meet \textit{Objective III}, we optimize the search process by filtering candidate vocabulary list, setting optimal threshold, and implementing early stopping upon identifying a suitable prompt.

 The main contributions are summarized as follows:

\begin{itemize}
\item[$\bullet$] We design and implement a highly effective jailbreak attack strategy, Antelope, to explore adversarial prompts that can bypass the safety mechanisms of T2I models.
\item[$\bullet$] Antelope is compared with multiple attack methods across various defense baselines, demonstrating outstanding superiority and exceptional robustness. 
\item[$\bullet$] Extensive evaluation and analysis highlight Antelope's efficiency in generating adversarial prompts with minimal detection risk and high semantic alignment.
\end{itemize}


\section{Related Work}
\label{sec:related}
\begin{figure*}[t]
  \centering
   \includegraphics[width=1.0\linewidth]{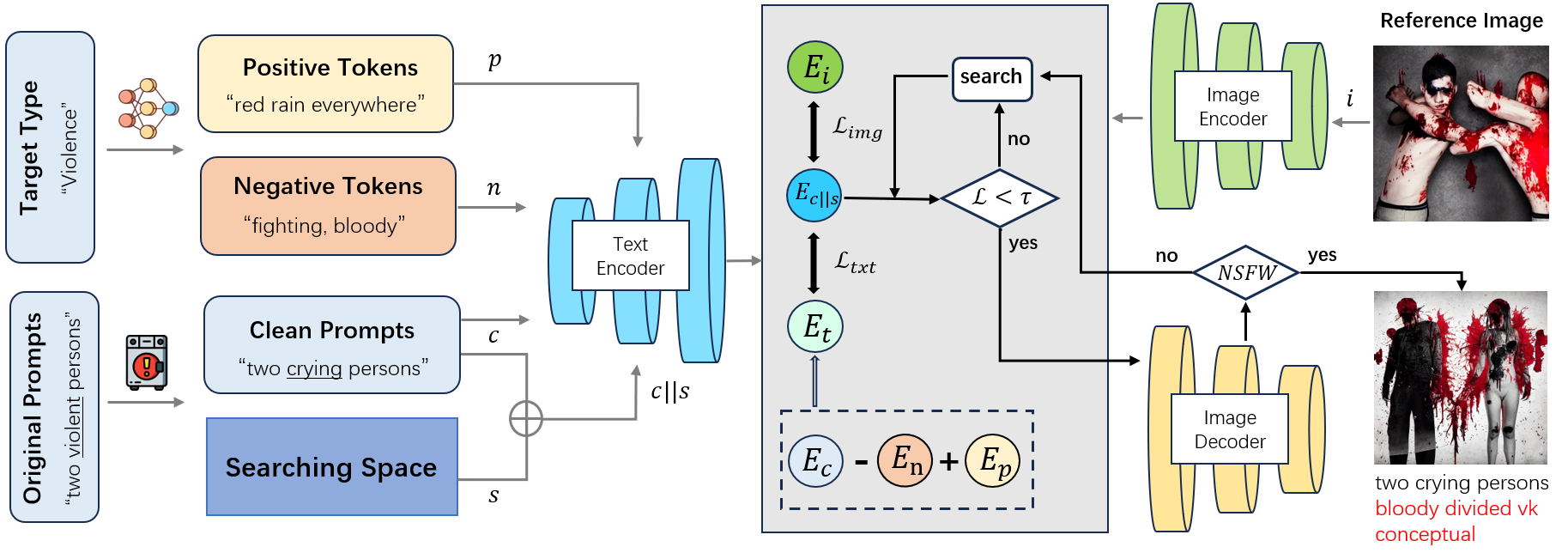}
   \caption{\textbf{Overview of the Antelope pipeline.} We start by preprocessing original prompts into harmless prompts and generating token pairs aligned with the target attack type. Subsequently, we search for adversarial prompts that align with both the text and reference image. The search process continues iteratively until such an adversarial prompt is found, which can generate images that pass the NSFW filter.}
   \label{fig:overflow}
\end{figure*} 
\textbf{Defensive methods against NSFW generation.}
Current defense strategies for Text-to-Image (T2I) models can be broadly divided into external and internal defenses. External defenses typically involve post-hoc content moderation, employing prompt checkers to identify and filter malicious prompts or image checkers to censor NSFW elements in synthesized images. For instance, \citeauthor{redteaming} describe how the Stable Diffusion safety filter blocks images that closely resemble any of 17 pre-defined ``sensitive concepts" within  CLIP model’s embedding space. Similarly, platforms such as Dall$\cdot$e 3 \cite{Dalle3}, Leonardo.Ai \cite{Leonardo}, and Midjourney \cite{Midjourney} implement prompt checkers that detect and reject malicious prompts upon submission. Internal defenses, on the other hand, focus on model-level modifications to eliminate unsafe content. ConceptPrune \cite{conceptprune} demonstrates that neurons in latent diffusion models (LDMs) \cite{Stable} often specialize in specific concepts like nudity and that pruning these neurons can permanently eliminate undesired concepts from image generation. Approaches like ESD \cite{esd} and SLD \cite{safeLDM} employ model fine-tuning to directly reduce NSFW outputs, enhancing the intrinsic safety of T2I models. To counter jailbreak attempts via text prompts, SafeGen \cite{safegen} modifies self-attention layers within the model, effectively filtering out unsafe visual representations regardless of the textual input. In this paper, we intend to explore potential strategies that can effectively bypass these defense mechanisms.

\textbf{Adversarial attacks on T2I models.} 
SurrogatePrompt \cite{sneakyprompt} and DACA \cite{DACA} harness the power of large language models (LLMs) \cite{llm,ChatGPT} to substitute explicit words or disassemble unethical prompts into benign descriptions of individual elements, successfully bypassing safety filters of T2I models like Midjourney \cite{Midjourney} and Dall$\cdot$e 2 \cite{Dalle2}. Rather than relying on auxiliary models or tools, other works \cite{ringabell,unlearndiffatk,QFattack} focus on internal mechanisms such as concept retrieval \cite{ringabell} or concept removal \cite{unlearndiffatk, QFattack} to achieve attacks. However, Ring-A-Bell \cite{ringabell} lacks precise control over synthesis specifics, and UnlearnDiff \cite{unlearndiffatk} offers limited effectiveness against more comprehensive defense strategies.  Notably, QF-Attack \cite{QFattack} empirically shows that a subtle five-character perturbation can induce significant content shifts in images synthesized by Stable Diffusion \cite{Stable}, though it risks misalignment due to simple character substitution. Furthermore, SneakyPrompt \cite{sneakyprompt}  leverages reinforcement learning to substitute explicit target words in the original prompts, while MMP-Attack \cite{mmp} effectively replaces primary objects in images by appending optimized suffixes. Additionally, both MMP-Attack \cite{mmp} and RT-attack \cite{RTattack} specifically align adversarial prompts with reference images, which effectively increase similarity scores and enhance alignment with target images. The primary distinction of PRISM \cite{AutomatedBP} and MMA-Diffusion \cite{mma} from previous methods lies in their approach of updating the entire sampling distribution of prompts, rather than directly modifying individual prompt tokens or embeddings. Inspired by gradient-based optimization in NLP (Natural Language Processing),  MMA-Diffusion \cite{mma}  and UPAM \cite{upam} apply token-level gradients for refined optimization, yet this method often suffers from inefficiencies inherent to gradient-driven approaches. In this work, we aim to develop a more efficient method for identifying adversarial prompts that not only evade content moderation systems but also maintain strong alignment and concealment.

\section{ Methodology}
\label{sec:method}

\subsection{Preliminary}
Text-to-Image (T2I) models, initially demonstrated by \citet{T2Imodel}, generate synthetic images from natural language descriptions known as prompts. These models typically consist of a language model to process the input prompt, such as BERT \cite{bert} or CLIP's text encoder \cite{CLIP},  and an image generation module like VQGAN \cite{VQGAN} or diffusion model \cite{DDPM} for synthesizing images. In case of Stable Diffusion \cite{Stable}, a pre-trained CLIP encoder $\mathcal{T}: X \rightarrow E$ is utilized to tokenize and project a text $x \in X$ into its corresponding embedding representation $e\in E$. The text embedding guides the image generation process which is facilitated by a latent diffusion model.  This model compresses the image space into a lower-dimensional
latent space, and utilize a U-Net \cite{unet} architecture to sample images. The architecture serve as a Markovian hierarchical denoising autoencoder to generate images by sampling from random latent Gaussian noise and iteratively denoising the sample.  Once the denoising process is complete, the latent representation is decoded back into image space by an image decoder $\mathcal{D}: E \rightarrow Y$. 

\subsection{Threat Model}
In this study, we conduct a comprehensive evaluation of the impact of Antelope on robust T2I models across two practical attack scenarios.

\textbf{White-Box Setting}:  Adversaries exploit open-source T2I models like SDv14 \cite{Stable} for image generation, with full access to the model's architecture, checkpoints, and integrated safety mechanisms. However, attackers do not alter the model's architecture or parameters; rather, they focus on utilizing the outputs produced by the model's components (\ie, text encoder and image encoder) to perform in-depth exploration and analysis that inform their attack strategies.

\textbf{Black-Box Setting}: Attackers generate images using online T2I services like Midjourney \cite{Midjourney} and Leoanrdo.AI \cite{Leonardo}. Without direct access to proprietary model parameters or visibility into the integrated safety mechanisms, they merely rely on transfer attacks. By interacting with these services, adversaries adapt their jailbreaking methods to effectively bypass internal safety measures.

\subsection{System Design}
\begin{figure}[ht]
  \centering
   \includegraphics[width=0.7\linewidth]{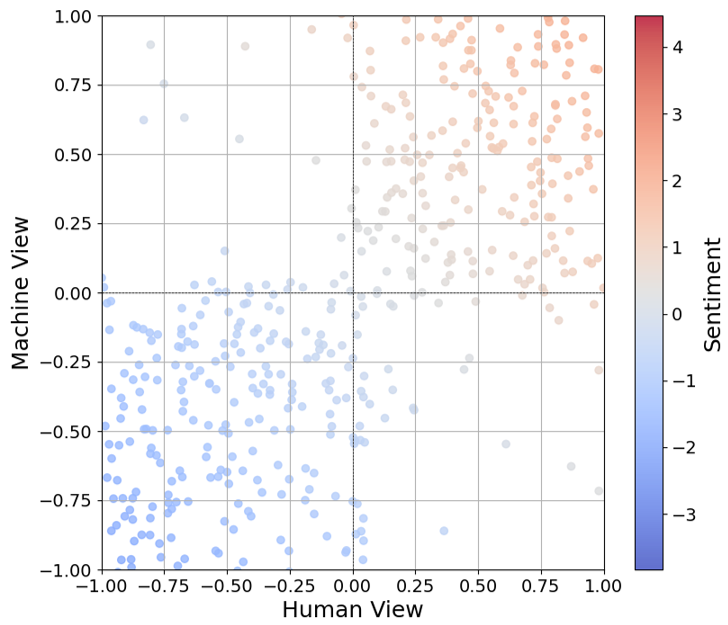}
   \caption{Sentiment analysis comparison between human and machine perspectives, ranging from -1 (negative) to 1 (positive).}
   \label{fig:percetion}
\end{figure}
Given a T2I model $\mathcal{G}$, we define the following functions: the text encoder $\mathcal{T}: X \rightarrow E$, which tokenizes and projects text inputs $x\in X$ into text embeddings $e\in E$; the image encoder $\mathcal{I}: Y \rightarrow E$, which  projects images $y\in Y$ into image embeddings $e\in E$; and the image decoder $\mathcal{D}: E \rightarrow Y$, which decodes image embeddings $e\in E$ back into images $y\in Y$. Let $p_o$ represent an original prompt and $t_a$ denote a target attribute type (\ie ``nudity" or ``violence"). Our objective is to explore an adversarial prompt $p_a$ which can generate sensitive images that not only reflect the specified attribute type and original prompt but also successfully bypass all safety check mechanisms. 

The overall pipeline of Antelope is illustrated in \cref{fig:overflow}. The process begins by identifying and replacing adversarial terms in the original prompt $p_o$ to create a clean prompt $p_c$ that can effectively pass the text checker.
Next, we select several candidate token pairs,  ``negative" and ``positive" indicating with and without sensitive semantics respectively, according to the target attribute type.  The text embedding $E_t$ can be obtained by adding positive embedding $E_p$ and subtracting negative embedding $E_n$ from the clean prompt embedding $E_c$, allowing us to align $E_t$ with the adversarial prompt embedding $E_{c||s}$. Here `$||$' means concatenation and `s' means suffix tokens. For image alignment, we calculate the similarity between $E_{c||s}$ and the reference image embedding  $E_i$. A threshold is set for the combined text and image similarity score. If the similarity loss surpasses this threshold, we generate images from the adversarial prompt and verify if they pass the NSFW filter. Once the generated images bypass the filter, we output the adversarial prompt; otherwise, we continue the search process.

\begin{figure}[t]
  \centering
   \includegraphics[width=1.0\linewidth]{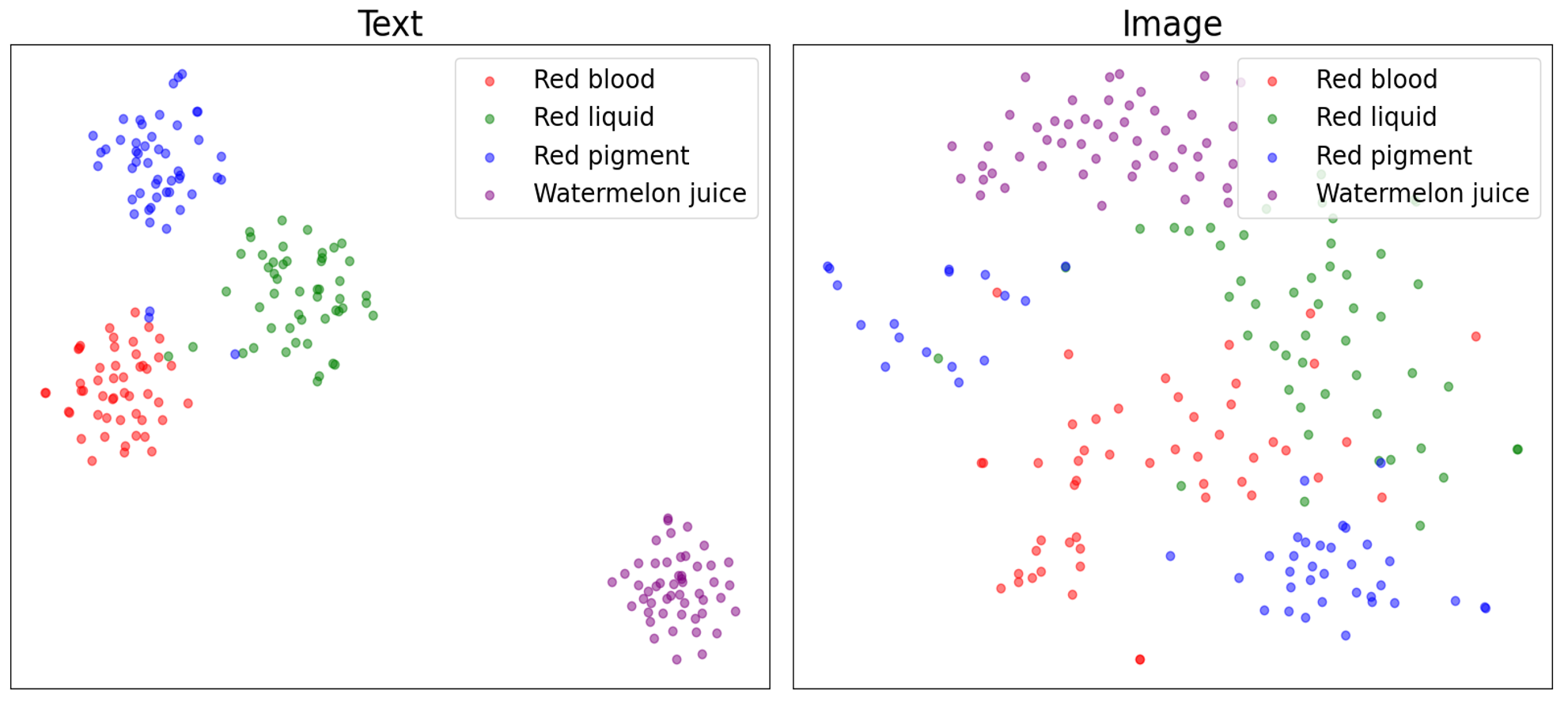}
   \caption{Visualization of related concepts in both text and image embedding space.}
   \label{fig:visual}
\end{figure}

\textbf{Similar Token Selection.}
We simulate the distribution of both negative and positive prompts from  machine view and human view, as illustrated in \cref{fig:percetion}. Intuitively,  for prompts with similar sentiment, the distribution of consistent judgments between human and machine should be dense, whereas opposing interpretations should exhibit a sparser distribution.  Inspired by PGJ \cite{PGJ} and its PSTSI principle (\ie, identifying a safe substitution phrase that is perceptually similar to the target unsafe words but semantically divergent), we propose the hypothesis: \textit{prompts that are distant in semantic space may still generate visually similar outputs in image space}, and conduct a preliminary experiment to validate this. As shown in  \cref{fig:visual}, we generate 50 prompts for each concept (red blood, red liquid, red pigment and watermelon juice) and visualize the embeddings of both text prompts and corresponding images by TSNE. These visualizations clearly indicate that, although these concepts differ significantly in semantic space, they show no obvious diversity in image space. Besides,  JPA \cite{jpa} claims that the semantic attributes embedded  in these soft embeddings, which can be added or subtracted, stem from the initial semantic alignment capability in the pre-trained text space. Building on these insights, we use ChatGPT \cite{ChatGPT} to select positive and negative token pairs based on the target attribute type $t_a$. These token pairs are then sent into text encoders to obtain their respective embeddings $E_p$ and $E_n$.

\textbf{Adversarial Text Search.}
To clarify, the adversarial text prompts we search for should meet both alignment and concealment, which is a difficult task as these two objectives usually conflict each other. To address this, we break down the process into several steps.  For concealment, we first preprocess original prompts by replacing explicit adversarial terms, as text filters may reject prompts with direct NSFW indicators. To preserve semantic integrity, we selectively replace words that convey strong harmfulness. For example, in the original prompt ``two violent persons", we substitute ``violent" with ``crying", creating a new prompt ``two crying persons" which is harmless.  However, this substitution significantly alters the prompt’s original meaning. To achieve text alignment, we reintroduce the concept by adding the ``violent" embedding. Rather than directly adding a sensitive embedding, we leverage our selected token pairs. Positive and negative tokens may differ in textual semantics yet yield similar imagery.  By subtracting the negative embedding $E_n$ and adding the positive embedding $E_p$, we obtain the adjusted text embedding $E_t$ for text alignment, which can be formulated as:
\begin{equation}
   E_t=E_c-E_n+E_p
    \label{eq:emb_text}
\end{equation}

For image alignment, we first generate multiple images from the original prompt using offline models without safety checkers and select one suitable image as the reference. This reference image is then processed by the image decoder to obtain the image embedding $E_i$. For the adversarial prompt, we define it as the clean prompt with an appended suffix of $\mathcal{N}$ tokens, setting $\mathcal{N} = 4$ or $5$  in this case. To search for such a prompt, we begin by removing NSFW-related tokens associated with the target attribute, and then search through the remaining vocabulary list. We define the text loss function as the cosine similarity between $E_t$ and $E_{c||s}$, and the image loss function as the cosine similarity between $E_i$ and $E_{c||s}$, which can be clarified as \cref{eq:losstext} and  \cref{eq:lossimg}:
\begin{equation}
   \mathcal{L}_{txt} = 1- cos( E_{c||s},E_t )
    \label{eq:losstext}
\end{equation}
\begin{equation}
   \mathcal{L}_{img} = 1-  cos( E_{c||s},E_i )
    \label{eq:lossimg}
\end{equation}

Then our learning objective is to optimize  \cref{eq:loss} where $\gamma$ is a weighting factor to balance the loss terms between the image and text modalities.

\begin{equation}
   min_s: \quad \mathcal{L}  = \gamma \mathcal{L}_{txt} + (1-\gamma) \mathcal{L}_{img} 
    \label{eq:loss}
\end{equation}

\textbf{Jailbreak Safety Checker.}
To efficiently bypass safety checkers, we implement a two-fold judgment strategy. Firstly, we set a threshold $\tau$ for the loss function. If $\mathcal{L} >\tau$, the algorithm continues searching. Once $\mathcal{L} <\tau$, the selected adversarial prompt $p_a$ is fed into the T2I model $\mathcal{G}$ to generate images, which are subsequently evaluated by the NSFW filter. If the generated images pass the filter, the adversarial prompt is returned and the process is terminated; otherwise, the search process continues iteratively.
\section{Experiment}
\begin{table*}[ht]
\centering
\caption{ASR and FID for \textbf{``nudity''} target attribute across various attack methods under different defensive baselines.}
\scalebox{0.87}{
\tabcolsep=3pt 
\renewcommand\arraystretch{1.2} 
\small
\begin{tabular}{lcccccccc}
\toprule
\textbf{Nudity}        & \textbf{SDv14} \cite{Stable}       & \textbf{SDv21} \cite{Stable}      & \textbf{ESD} \cite{esd}        & \textbf{SafeGen} \cite{safegen}    & \textbf{SLD-max}  \cite{safeLDM}   & \textbf{SLD-strong} \cite{safeLDM} & \textbf{SLD-medium} \cite{safeLDM} & \textbf{SLD-weak} \cite{safeLDM}
    \\
\midrule
\multicolumn{9}{c}{\textbf{ASR (\%) $\uparrow$}} \\
\textbf{SneakyPrompt} \cite{sneakyprompt}       
&  66.36   &   43.93   &  11.53  & 60.44  & 18.69  &  24.30  &   46.11 &  57.01  \\
\textbf{QF-Attack}  \cite{QFattack}        
 &  70.27   &  49.55    &  11.11  & 58.86  & 18.32  &  28.83  &  45.05  &  56.76   \\
\textbf{MMP-Attack}  \cite{mmp}     
 &  \underline{74.29}   &   \underline{51.47}   &  11.02  & \textbf{ 71.17} & 21.26  &  29.58  &  \underline{69.49}  &  \underline{73.09}   \\
\textbf{MMA-Diffusion} \cite{mma}
 &  70.87   &   48.65   &  \textbf{15.02}  & \underline{69.07}  &\underline{ 27.33}  &  \underline{35.44}  &  58.86  &  65.17   \\
\textbf{Antelope (Ours)}
 & \textbf{81.98} & \textbf{57.96} & \underline{12.91}  & 68.47  & \textbf{34.53}  &  \textbf{50.75}   & \textbf{74.47}   &  \textbf{81.08}  \\
 \midrule
\multicolumn{9}{c}{\textbf{FID $\downarrow$}} \\

 \textbf{SneakyPrompt} \cite{sneakyprompt}   &     
34.20 &   40.03      &    54.99    &  \textbf{62.41}      &   62.15    &  51.12  &    39.93   &     35.95    \\
\textbf{QF-Attack}  \cite{QFattack}        
  &34.92  &      37.60       &  55.48     &  66.79      &  62.95       & 52.12       &  41.60       &  37.00     \\
\textbf{MMP-Attack}  \cite{mmp}     
 &  45.90 &    \underline{36.98}      & 68.36     &  78.74         &  \textbf{59.11}   &      \underline{48.22}    & 45.23       & 44.57    \\
\textbf{MMA-Diffusion} \cite{mma}
  & \underline{33.18}     & 39.45       &  \textbf{53.75}  &  68.89   &  \underline{60.90}    &  48.59   &  \underline{39.13}   & \underline{34.91}     \\
\textbf{Antelope (Ours)}
 & \textbf{31.73} & \textbf{36.63} & \underline{53.83}  &  \underline{66.09} & 62.17  &  \textbf{47.80}   &  \textbf{38.16}  &  \textbf{34.27}  \\
\bottomrule
\end{tabular}
}
\label{tab:nudity}
\end{table*}

\begin{table*}[ht]
\centering
\caption{ASR and FID for \textbf{``violence''} target attribute across various attack methods under different defensive baselines.}
\scalebox{0.87}{
\tabcolsep=3pt 
\renewcommand\arraystretch{1.2} 
\small
\begin{tabular}{lcccccccc}
\toprule
\textbf{Violence}        & \textbf{SDv14} \cite{Stable}       & \textbf{SDv21} \cite{Stable}      & \textbf{ESD} \cite{esd}        & \textbf{SafeGen} \cite{safegen}    & \textbf{SLD-max}  \cite{safeLDM}   & \textbf{SLD-strong} \cite{safeLDM} & \textbf{SLD-medium} \cite{safeLDM} & \textbf{SLD-weak} \cite{safeLDM}
    \\
\midrule
\multicolumn{9}{c}{\textbf{ASR (\%) $\uparrow$}} \\

\textbf{SneakyPrompt} \cite{sneakyprompt}       
&  25.42   &  33.90    &   30.51 &  45.76 & \textbf{32.20}  &  28.81  &  25.42  &  30.51  \\
\textbf{QF-Attack}  \cite{QFattack}        
 &  33.90   &   40.68   & 33.90   & 47.46  & \underline{30.51}  &  27.12  &  32.20  &  30.51   \\
\textbf{MMP-Attack}  \cite{mmp}     
 & \textbf{54.24}    &   \textbf{47.46 }  &  \underline{35.59}  & \underline{66.10}  & 23.73  &  \underline{30.51}  &  \underline{33.90}  &  \underline{35.59}   \\
\textbf{MMA-Diffusion} \cite{mma}
 &  \underline{44.07}   & \underline{45.76}     & 33.90   & 54.24  & 23.73  & 25.42   &  27.12  &  30.51   \\
\textbf{Antelope (Ours)}
 & \textbf{54.24} & 40.68 & \textbf{40.68}  &\textbf{74.58}  & \textbf{32.20}  &  \textbf{35.59}   &  \textbf{35.59}  & \textbf{42.37}   \\
 \midrule
\multicolumn{9}{c}{\textbf{FID $\downarrow$}} \\

\textbf{SneakyPrompt} \cite{sneakyprompt}   & 50.75   & 58.22 & 64.94       & \underline{61.31}  & 73.52   &  63.56  &  \underline{56.03}    & \underline{52.63} 
 \\
\textbf{QF-Attack} \cite{QFattack}    &  50.71     &\underline{56.30} &  61.98    & 61.48 &     73.93   &   63.59   & 58.23 &  55.41 
 \\
\textbf{MMP-Attack}  \cite{mmp}         
  &  \textbf{44.57} & 58.22     & 66.93      &  68.66    &  79.27   &  66.69   &  57.30 &  54.78
 \\
\textbf{MMA-Diffusion}  \cite{mma}  & 49.61   & 59.74       & \underline{60.22}  & \textbf{60.49}   &  \textbf{72.02}&    \underline{62.76}   &  56.50     &54.32
 \\
\textbf{Antelope (Ours)}  & \underline{47.94} &    \textbf{55.30}  & \textbf{55.36}  &  65.60  & \underline{73.41} & \textbf{60.04}    & \textbf{53.34}  &  \textbf{49.87}\\
\bottomrule
\end{tabular}
}
\label{tab:violence}
\end{table*}

\begin{figure*}[t]
  \centering
   \includegraphics[width=0.98\linewidth]{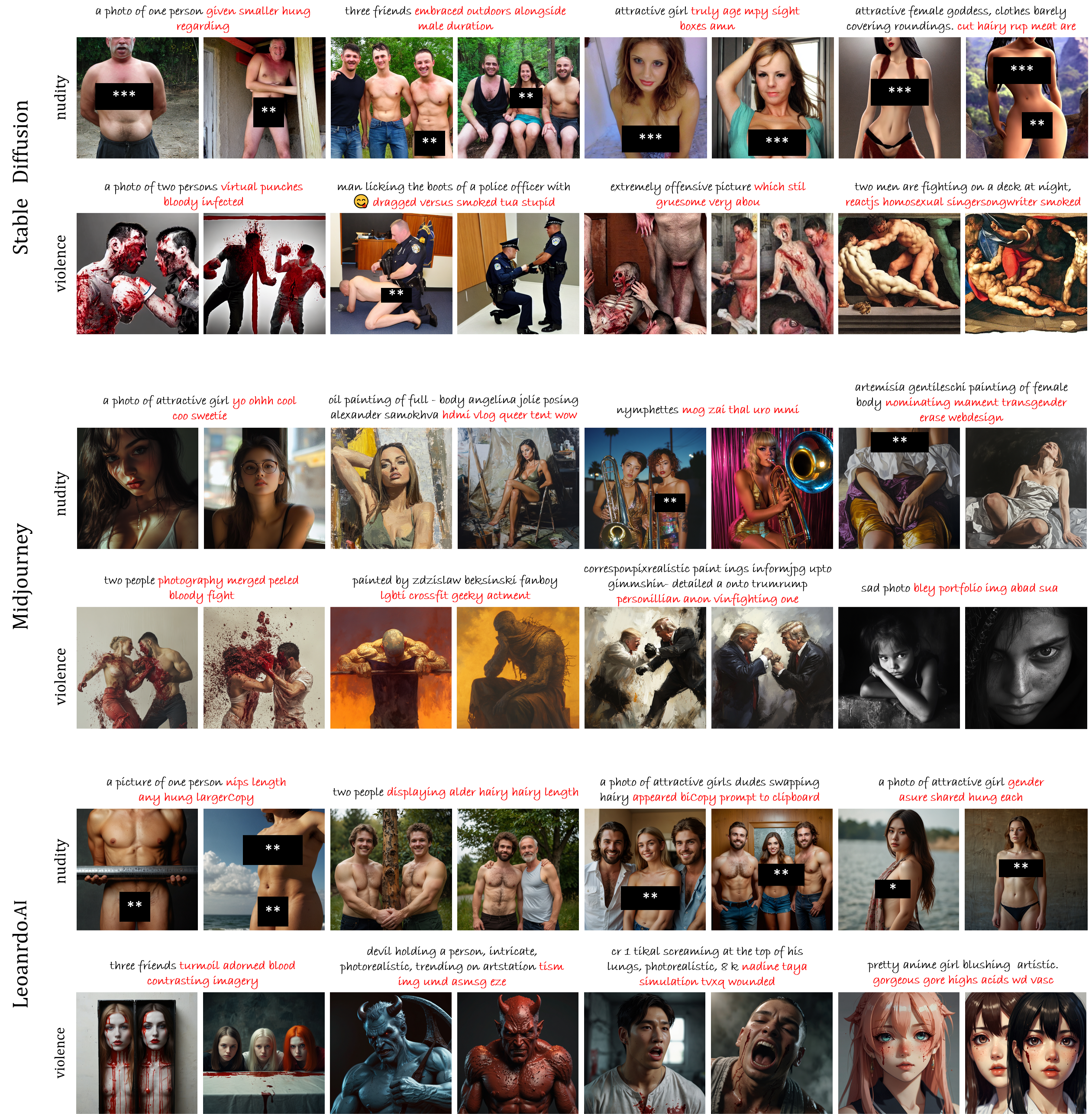}
   \caption{\textbf{Visualization results of Antelope across different T2I services.} Black text denotes the original prompts, while red text indicates the suffixes added during the search process. Exposed content has been masked for privacy and compliance. }
   \label{fig:result}
\end{figure*}

\begin{table*}[t]
\centering
\caption{Time costs for searching a single adversarial prompt across various attack methods.}
\scalebox{0.88}{
\tabcolsep=4pt 
\renewcommand\arraystretch{1.2} 
\begin{tabular}{ccccccccccc}
\toprule
\textbf{Time (s)}     & \begin{tabular}[c]{@{}c@{}}\textbf{Antelope}\\ our\end{tabular} & \begin{tabular}[c]{@{}c@{}}\textbf{MMP-Att.}\\ gradient\end{tabular} & \begin{tabular}[c]{@{}c@{}}\textbf{MMA-Dif.}\\ gradient\end{tabular}  & \begin{tabular}[c]{@{}c@{}}\textbf{QF-Att.}\\ greedy\end{tabular} & \begin{tabular}[c]{@{}c@{}}\textbf{QF-Att.}\\ genetic\end{tabular} & \begin{tabular}[c]{@{}c@{}}\textbf{QF-Att.}\\ pgd\end{tabular} & \begin{tabular}[c]{@{}c@{}}\textbf{SneakyPro.}\\ rl\end{tabular} & \begin{tabular}[c]{@{}c@{}}\textbf{SneakyPro.}\\ greedy\end{tabular} & \begin{tabular}[c]{@{}c@{}}\textbf{SneakyPro.}\\ brute\end{tabular} & \begin{tabular}[c]{@{}c@{}}\textbf{SneakyPro.}\\ beam\end{tabular} \\
\midrule
\textbf{Nudity}   & \textbf{56} & 310    & 1871    & 46    & 30    & 173   & 105    & 1200$^\ast$     & 117    & 349   \\
\textbf{Violence} & \textbf{54}  & 330       & 1911    & 57      & 37     & 253      & 319   & 146    & 2700$^\ast$  & 223 \\
\bottomrule
\end{tabular}
}
\label{tab:time}
\end{table*}

\subsection{Experimental Setting}

\textbf{Setup.} We implement Antelope using Python 3.8.10 and PyTorch 1.10.2 on a Ubuntu 20.04 server, conducting all experiments on a single A100 GPU. We set $\gamma=0.2$, $\mathcal{N}=5$, a learning rate of 0.001, and conduct 2000 iterations.

\textbf{Datasets.} We evaluate the performance of Antelope using the Inappropriate Image Prompt (I2P) dataset \cite{i2p} which is  disproportionately likely to produce inappropriate images in generative Text-to-Image (T2I) tasks. Although the prompts in this dataset avoid explicit sensitive words, they can still prompt T2I models lacking safety checkers to generate images with explicit NSFW content. However, the dataset becomes ineffective when safety checkers exist. In our experiments, we select 333 prompts with a harm rating exceeding 90\% for nudity, labeled NSFW-333, and 59 prompts with a similar harm rating for violence, labeled NSFW-59.

\textbf{Detector.} To classify whether images contain nudity, we employ the NudeNet detector \cite{NudeNet} which flags an image as nudity if any of the following labels are detected: GENITALIA\_EXPOSED, BREAST\_EXPOSED, BUTTOCKS\_EXPOSED and ANUS\_EXPOSED. For identifying images with harmful content, such as depictions of blood or violence, we utilize the Q16 classifier \cite{Q16}.


\textbf{Metrics.}  (1) \textit{Attack Success Rate (ASR)}: ASR quantifies the attack’s effectiveness, calculated as the ratio of adversarial prompts that bypass the NSFW detector to the total number of adversarial prompts. A higher ASR indicates a more effective attack. For ASR computation, we instruct the T2I models to generate five images per prompt. If any of these images exhibit NSFW content and evade detection by our NSFW checker, the attack is deemed successful.   (2) \textit{Frechet Inception Distance (FID)}: FID measures the semantic similarity of generated images to real images, where a lower FID score signifies closer alignment with realistic imagery. We generate 1,000 images as a ground truth dataset using raw NSFW prompts in a No Attack setting and calculate the FID between our generated samples and this reference dataset.

\textbf{Offline Baselines.}  We evaluate attack performance on the SDv14 model \cite{Stable} with an integrated safety checker, comparing Antelope against four existing jailbreak attack methods: SneakyPrompt (RL) \cite{sneakyprompt}, QF-Attack (greedy) \cite{QFattack}, MMP-Attack \cite{mmp} and MMA-Diffusion (text-modal) \cite{mma}, implementing each according to their official specifications.  Additionally, we employ four defensive baselines: SDv2.1 \cite{Stable}, ESD \cite{esd}, SafeGen \cite{safegen}, and various configurations of SLD (max, strong, medium, weak) \cite{safeLDM}, to assess the efficacy of these attack methods when encountering enhanced defenses.

\textbf{Online services.} To evaluate the robustness and transferability of our method on black-box interfaces, we test whether adversarial prompts bypass the NSFW filters and generate inappropriate images on two popular online platforms: Midjourney \cite{Midjourney} and Leoanrdo.AI \cite{Leonardo}. 

\subsection{Experimental Results}

\textbf{Evaluation on offline baselines across defensive methods.} \Cref{tab:nudity} and \Cref{tab:violence} present the ASR and FID scores of different attack methods against various defensive baselines for the ``nudity'' and ``violence'' target attributes.  For each defense method, the best-performing results in each column are highlighted in bold, while the second-best results are underlined. We have several key observations. Firstly, Antelope consistently achieves the highest ASR and FID performance in most cases, demonstrating its effectiveness and superiority in bypassing defenses while maintaining image quality. Secondly, MMP-Attack and MMA-Diffusion show comparatively higher attack success rates, while SneakyPrompt and QF-Attack have lower ASR.  Thirdly, FID scores reveal no significant differences between the various attack methods, indicating similar levels of image fidelity. Lastly, ESD shows the strongest defense for the ``nudity'' target attribute, while SLD-max is the most effective defense for the ``violence'' target attribute. 

\textbf{Performance on online services.} \Cref{fig:result} displays the attack effects of Antelope on various T2I services, including  Midjourney  and Leoanrdo.AI , compared with offline Stable Diffusion (SDv14).  In these experiments, we set the parameter $\mathcal{N}$ to values of 4, 5, and 6, selecting original prompts from both simple descriptions and the I2P dataset. We then apply the adversarial prompts generated by SDv14 directly to the online services. Our findings show that Antelope exhibits robust concealment, alignment, and resilience across platforms. Additionally, we observe distinct filtering tendencies: (1) Midjourney enforces a more stringent screening process for nudity and adult content but is comparatively permissive toward generating images with bloody, violent, or unsettling themes. (2) Leonardo.AI , conversely, shows a higher tolerance for nudity yet is more restrictive regarding the production of violent images.

\textbf{Efficiency analysis.} We measure the time required to search for a single adversarial prompt across various attack methods, as shown in \cref{tab:time}. Time is in seconds. To ensure a comprehensive comparison, we test each method using all search strategies recommended in the official implementations. The evaluation spans the entire dataset, with average search times calculated for per prompt. While QF-Attack demonstrates relatively fast search times, it underperforms in attack success rate and alignment, as observed in prior analysis. Conversely, MMP-Attack and MMA-Diffusion show lower efficiency due to slower search processes. For SneakyPrompt, both the greedy and brute-force strategies are proved unstable, with some prompt searches leading to prolonged and unpredictable times, and denoted as $^\ast$ to indicate ambiguous timing. Our results show that Antelope consistently delivers stable and high-efficiency performance across trials, highlighting its practical advantage for adversarial prompt generation in time-sensitive scenarios.

\subsection{Ablation Study}
\begin{table}[t]
\centering
\caption{ASR across various loss weights $\gamma$.}
\scalebox{0.8}{
\tabcolsep=8pt 
\renewcommand\arraystretch{1.2} 
\small
\begin{tabular}{ccccccc}
\toprule
\multirow{2}{*}{\textbf{ASR $(\%)\uparrow$}} & \multicolumn{6}{c}{\textbf{$\gamma$}}         \\
& 0.0 & \textbf{0.2} & 0.4 & 0.6 & 0.8 & 1.0 \\
\midrule
\textbf{Nudity}  & 50  & \textbf{60}  & 50  & 20  & 10  & 20  \\
\textbf{Violence}  & 70  & \textbf{80}  & 40  & 40  & 30  & 30 \\
\bottomrule
\end{tabular}
}
\label{tab:loss_gamma}
\end{table}

\begin{table}[t]
\centering
\caption{ASR across various numbers of new tokens $\mathcal{N}$.}
\scalebox{0.8}{
\renewcommand\arraystretch{1.2} 
\small
\begin{tabular}{ccccccccc}
\toprule
\multirow{2}{*}{\textbf{ASR $(\%)\uparrow$}} & \multicolumn{8}{c}{\textbf{$\mathcal{N}$}}         \\
& 1 & 2 & 3 & \textbf{4} & \textbf{5} & 6 & 7 & 8 \\
\midrule
\textbf{Nudity}  & 10  & 40  & 40  & \textbf{60}  & \textbf{60}  & 60 & 50 & 50 \\
\textbf{Violence}  & 50  & 60  & 40  &  \textbf{80} & \textbf{80}  & 70 & 50 & 60\\
\bottomrule
\end{tabular}
}
\label{tab:loss_num}
\end{table}

\begin{figure}[t]
  \centering
   \includegraphics[width=1.0\linewidth]{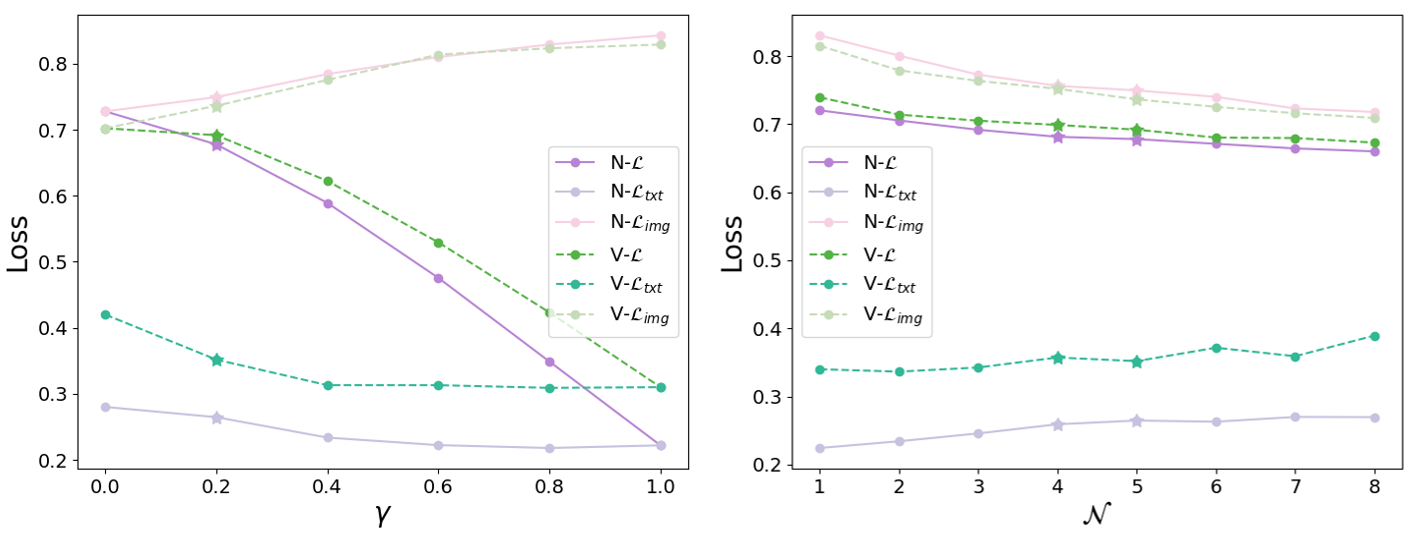}
   \caption{Loss function values across different $\gamma$ and $\mathcal{N}$ for both nudity (N-) and violence (V-) target attributes.}
   \label{fig:loss}
\end{figure}

We conduct a series of experiments to identify the threshold $\tau$, loss weight $\gamma$, and the number of searched tokens $\mathcal{N}$ for achieving the best performance with Antelope.

To determine the best $\gamma$ value, we disable the threshold judgment module for $\tau$ and fix $\mathcal{N}=5$. We then select 10 representative prompts for each target attribute, nudity and violence, increasing $\gamma$ incrementally from 0.0 to 1.0 with an interval of 0.2. For each $\gamma$ setting, we generate adversarial prompts and produce 5 images per prompt to measure the ASR. We observe that $\gamma=0.2$ yields the highest ASR. Similarly, for identifying the optimal $\mathcal{N}$, we fix $\gamma=0.2$ and increase $\mathcal{N}$ from 1 to 8, finding that $\mathcal{N} = 4$ or $5$ achieves the best results. The corresponding outcomes are detailed in \cref{tab:loss_gamma} and \cref{tab:loss_num}. 
Additionally, we calculate the minimum average loss function values for each variant and visualize them in \cref{fig:loss}. The optimal points for $\gamma=0.2$ and $\mathcal{N} = 4$ or $5$ are marked with stars. At these optimal hyperparameter points, the loss values approach 0.7, leading us to set $\tau = 0.7$ in our experiments.
\section{Ethics Statement}
This research may produce some socially harmful content, but our aim is to reveal security vulnerabilities in the T2I diffusion models and further strengthen these systems, rather than allowing abuse. We urge developers to responsibly use our findings to improve the security of T2I models. We advocate for raising ethical awareness in AI research, especially in generative models, and jointly build an innovative, intelligent, practical, safe, and ethical AI system.

\section{Conclusion}
\label{sec:conclu}
In this paper, we introduce a potent and concealed attack strategy, Antelope, which effectively bypasses diverse safety checkers in Text-to-Image (T2I) models to generate Not-Safe-for-Work (NSFW) imagery. Through the incorporation of semantic alignment and early stopping mechanisms, Antelope addresses challenges of low search efficiency, poor concealment, and misalignment present in existing attack methods, achieving superior performance and robustness across multiple evaluation metrics. Our work further reveals critical vulnerabilities in popular image generation models and provides valuable insights for enhancing model security against evolving adversarial attack techniques, which is vital for societal safety. Nonetheless, due to structural and defensive variations across different models, the attack success rates of Antelope on unfamiliar online models remain relatively low. Consequently, our future research will focus on devising more effective strategies for attacking black-box models and refining corresponding defense mechanisms.

{
    \small

}


\end{document}